%% file: main.tex
\pdfoutput=1 % for arxiv
\documentclass[]{spie}  %>>> use for US letter paper
\newcommand{\bicep}{B\textsc{icep}}
\newcommand{\keck}{\textit{Keck}}
\newcommand{\arry}{\textit{Array}}
\newcommand{\ukrts}{$\mu$K$\sqrt{\textrm{s}}$}
\newcommand{\p}{\phantom}
\newcommand{\tbf}{\textbf}
 
\usepackage{amsmath,amsfonts,amssymb}
\usepackage{graphicx}
\usepackage[colorlinks=true, allcolors=blue]{hyperref}
\usepackage{siunitx}
\usepackage{booktabs}
\usepackage{caption} 
\usepackage{color}
\usepackage{tablefootnote}

\title{BICEP3 performance overview and planned Keck Array upgrade}

\include{./authors}
\authorinfo{Corresponding author: J. A. Grayson, jgrayson@stanford.edu}

\begin{document} 
\maketitle

\begin{abstract}
  \bicep3 is a \SI{520}{mm} aperture, compact two-lens refractor designed to observe the polarization of the cosmic microwave background (CMB) at \SI{95}{GHz}. 
Its focal plane consists of modularized tiles of antenna-coupled transition edge sensors (TESs), similar to those used in \bicep2 and the \keck\,\arry. 
The increased per-receiver optical throughput compared to \bicep2/\keck\,\arry, due to both its faster $f/1.7$ optics and the larger aperture, more than doubles the combined mapping speed of the \bicep/\keck\,program. 
The \bicep3 receiver was recently upgraded to a full complement of 20 tiles of detectors (2560 TESs) and is now beginning its second year of observation (and first science season) at the South Pole. 
We report on its current performance and observing plans. Given its high per-receiver throughput while maintaining the advantages of a compact design, \bicep3-class receivers are ideally suited as building blocks for a 3rd-generation CMB experiment, consisting of multiple receivers spanning \SI{35}{GHz} to \SI{270}{GHz} with total detector count in the tens of thousands.
We present plans for such an array, the new ``\bicep\,Array'' that will replace the \keck\,\arry\ at the South Pole, including design optimization, frequency coverage, and deployment/observing strategies.
\end{abstract}

% Include a list of keywords after the abstract (up to 8). Will be searchable in SPIE database

% Body of text included in separate file
\include{./body}

% \appendix    %>>>> this command starts appendixes
% \section{MISCELLANEOUS FORMATTING DETAILS}
% \label{sec:misc}

% Acknowledgments
\acknowledgments
 
This work is supported by the National Science Foundation (grant nos. 1313158, 1313010, 1313062, 1313287, 1056465, 0960243), the SLAC Laboratory Directed Research and Development Fund, the Canada Foundation for Innovation, Science and Technology Facilities Council Consolidated Grant (ST/K000926/1), and the British Columbia Development Fund. The development of detector technology was supported by the JPL Research and Technology Development Fund and grants 06-ARPA206-0040, 10-SAT10-0017, and 12-SAT12-0031 from the NASA APRA and SAT programs. The development and testing of detector modules was supported by the Gordon and Betty Moore Foundation.

% References
\bibliography{thesis} % bibliography data in report.bib
\bibliographystyle{spiebib} % makes bibtex use spiebib.bst

\end{document}

%% file: authors.tex
\author[a,b]{J. A. Grayson}
\author[c]{P. A. R. Ade}
\author[b,a]{Z. Ahmed}
\author[d]{K. D. Alexander}
\author[e]{M. Amiri}
\author[d]{D. Barkats}
\author[f,g]{S. J. Benton}
\author[d]{C. A. Bischoff}
\author[h,i]{J. J. Bock}
\author[d]{H. Boenish}
\author[d]{R. Bowens-Rubin}
\author[d]{I. Buder}
\author[j]{E. Bullock}
\author[d,k]{V. Buza}
\author[d]{J. Connors}
\author[h,l,m]{J. P. Filippini}
\author[n]{S. Fliescher}
\author[e]{M. Halpern}
\author[d]{S. Harrison}
\author[o]{G. C. Hilton}
\author[h]{V. V. Hristov}
\author[h]{H. Hui}
\author[a,b,o]{K. D. Irwin}
\author[a,b]{J. Kang}
\author[d]{K. S. Karkare}
\author[a,b]{E. Karpel}
\author[h]{S. Kefeli}
\author[a,b]{S. A. Kernasovskiy}
\author[d,k]{J. M. Kovac}
\author[a,b]{C. L. Kuo}
\author[p]{E. M. Leitch}
\author[h]{M. Lueker}
\author[i]{K. G. Megerian}
\author[a,b]{V. Monticue}
\author[a,b]{T. Namikawa}
\author[f,q]{C. B. Netterfield}
\author[i]{H. T. Nguyen}
\author[h,i]{R. O'Brient}
\author[a,b]{R. W. Ogburn IV}
\author[n,j]{C. Pryke}
\author[o]{C. D. Reintsema}
\author[d]{S. Richter}
\author[n]{R. Schwarz}
\author[d]{C. Sorensen}
\author[n,p]{C. D. Sheehy}
\author[h,i]{Z. K. Staniszewski}
\author[h]{B. Steinbach}
\author[h,r]{G. P. Teply}
\author[a,b]{K. L. Thompson}
\author[a,b]{J. E. Tolan}
\author[c]{C. Tucker}
\author[i]{A. D. Turner}
\author[d,s,p]{A. G. Vieregg}
\author[a,b]{A. Wandui}
\author[i]{A. C. Weber}
\author[e]{D. V. Wiebe}
\author[n]{J. Willmert}
\author[t,a,b]{W. L. K. Wu}
\author[a,b]{K. W. Yoon}

\affil[a]{Department of Physics, Stanford University, Stanford, CA 94305, USA}
\affil[b]{Kavli Institute for Particle Astrophysics and Cosmology, SLAC National Accelerator Laboratory, Menlo Park, CA 94025, USA}
\affil[c]{School of Physics and Astronomy, Cardiff University, Cardiff, CF24~3AA, United Kingdom}
\affil[d]{Harvard-Smithsonian Center for Astrophysics, Cambridge, MA 02138, USA}
\affil[e]{Department of Physics and Astronomy, University of British Columbia, Vancouver, BC, V6T~1Z1, Canada}
\affil[f]{Department of Physics, University of Toronto, Toronto, ON, M5S 1A7, Canada}
\affil[g]{Department of Physics, Princeton University, Princeton, NJ 08544, USA}
\affil[h]{Department of Physics, California Institute of Technology, Pasadena, CA 91125, USA}
\affil[i]{Jet Propulsion Laboratory, Pasadena, CA 91109, USA}
\affil[j]{Minnesota Institute for Astrophysics, University of Minnesota, Minneapolis, MN 55455, USA}
\affil[k]{Department of Physics, Harvard University, Cambridge, MA 02138, USA}
\affil[l]{Department of Physics, University of Illinois at Urbana-Champaign, Urbana, IL 61801, USA}
\affil[m]{Department of Astronomy, University of Illinois at Urbana-Champaign, Urbana, IL 61801, USA}
\affil[n]{School of Physics and Astronomy, University of Minnesota, Minneapolis, MN 55455, USA}
\affil[o]{National Institute of Standards and Technology, Boulder, CO 80305, USA}
\affil[p]{Kavli Institute for Cosmological Physics, University of Chicago, Chicago, IL 60637, USA}
\affil[q]{Canadian Institute for Advanced Research, Toronto, ON, M5G~1Z8, Canada}
\affil[r]{Department of Physics, University of California at San Diego, La Jolla, CA 92093, USA}
\affil[s]{Department of Physics, Enrico Fermi Institute, University of Chicago, Chicago, IL 60637, USA}
\affil[t]{Department of Physics, University of California, Berkeley, CA 94720, USA}

%% file: body.tex
\section{Introduction}
\label{sec:intro}

The precise measurement of the polarization of the Cosmic Microwave Background (CMB) is an active area of experimental physics, motivating a variety of current and upcoming polarimeter projects.
% IDB-D: supporting --> motivating
The CMB temperature anisotropy, measured to extreme precision over the last two decades, has provided evidence for the standard model of Big Bang cosmology and hints at an inflationary period. In the same way, the CMB polarization, and in particular the curl `B-mode' pattern, contains a wealth of cosmological information on many scales and continues to be targeted by experimentalists. At arcminute scales on the sky, the dominant B-mode signal is primarily due to the gravitational lensing of the much brighter, curl-free `E-mode' signal. This lensing B-mode signal probes the large scale structure of the universe that sits between the CMB surface and ourselves and has recently been detected by P\textsc{olarbear}\cite{pbear2014}, SPTpol\cite{sptpol2015}, and \bicep2/\keck\,\arry\cite{BKI,BKVI}. 

At larger, degree-scales on the sky, the B-mode signal due to inflationary gravitational waves (IGW) is predicted to peak with an as-yet-unmeasured amplitude. 
% KW-D: is theorized to peak with as yet unmeasured amplitude. --> is predicted to peak with an as-yet-unmeasured amplitude.
During inflation, quantum fluctuations are expected to produce both density and gravitational wave perturbations, the latter of which imprints a B-mode pattern on the CMB. The ratio of the amplitude of the primordial tensor spectrum to the scalar spectrum, parametrized as $r$, 
% ZA-D: density spectra --> scalar spectra
is a measurable quantity that will constrain inflationary models.
% ZA-D: will shed powerful light on the numerous inflationary models. 
% --> is a measurable qty that will constrain inflationary models.
% KW-D: and the ratio r of the amplitude of the gravitational to scalar spectra is a measurable quantity that will constrain inflationary models --> and the ratio of the amplitude of the primordial tensor spectrum to the scalar spectrum, parameterized as r, is a measurable quantity that will constrain inflationary models 
 To this end, experiments aim to continually lower the upper-limit on, and eventually detect, the IGW B-mode signal and the resulting $r$ parameter, a recent summary of which is given in Ref.~\citenum{kam2016}. 

The \bicep/\keck\ series of South Pole-based CMB polarimeters has focused efforts on this IGW B-mode signal, taking the unique approach of using compact, refracting telescope receivers that only resolve the larger degree-scale features necessary. This approach precludes science at the smaller scales but has several advantages for IGW signal sensitivity.
The first experiment, \bicep1, observed for three seasons from 2006--2008 using 98 feedhorn-coupled bolometers (split primarily between 100 and \SI{150}{GHz} band centers) and found a 95\% confidence upper limit of $r<0.70$\cite{barkats2014}. 
\bicep2 followed with three observing seasons from 2010--2012, using 500 antenna-coupled transition edge sensor (TES) bolometers centered at 150 GHz. It detected on-sky B-mode signal,
% ZA-D: This number looks wrong. r=0.2 at >5 sigma.
% sigma = 0.1 is from BA proposal, and B1xB2 result in B2 paper
later found to be consistent with galactic dust foregrounds, for a combined limit $r<0.12$ at 95\% confidence with \keck\,\arry\ and \textit{Planck} data\cite{BKI,bkp2015}.
While \bicep2 was still operating, \keck\,\arry\ began observing on a nearby telescope mount, with first science observations in 2012.
\keck\,\arry\ is made up of 5 \bicep2-class receivers sharing a common boresight. It is currently in its fifth observing season and has now included receivers centered on multiple frequencies (95, 150, and \SI{220}{GHz}) in order to help distinguish polarized galactic foregrounds from the IGW signal. The latest results give $r<0.09$ at 95\% confidence\cite{BKVI}, or $r<0.07$ when combined with external datasets.
% ZA-D: I suggest quoting 95%CL limits for results. sigma is a sensitivity.

\bicep3\ replaces \bicep2\ and is now in its second observing season (and first science season) during the 2016 austral winter, with a full complement of 2560 TES bolometers at 95 GHz. 
With its larger aperture, faster optics, and higher detector count,
% ZA-D: With its larger aperture, faster optics, and resulting high detector count, 
% --> larger aperture, faster optics, high detector count compared to BICEP2
\bicep3\ acts as a pathfinder receiver for the 3rd-generation of the \bicep/\keck\ series. Its relatively compact design with on-axis optics again lends itself to multi-frequency observations, in the same way that \bicep2\ was evolved into the \keck\,\arry. Such an expansion to an array is highly motivated by the continued need for increasing observation depth combined with discrimination of galactic foregrounds, and is easily realized by upgrading the \keck\,\arry\ with \bicep3-class receivers.

In the following sections, these proceedings first give an overview of the \bicep3 instrument (Sec.~\ref{sec:inst}) and its South Pole observing location and strategy (Sec.~\ref{sec:obs}). It then summarizes \bicep3's current, second-season status and performance (Sec.~\ref{sec:perf}) before finally outlining plans for a future multi-frequency array of \bicep3-class receivers (Sec.~\ref{sec:array}).

\section{Instrument Overview}
\label{sec:inst}

Like its \bicep/\keck\ predecessors described in Sec.~\ref{sec:intro}, \bicep3\ is composed of a single cryostat, also called the receiver, which contains all of the detectors and optical components required for its measurement of the CMB. This compact design, which keeps all optics at cryogenic temperatures, maintains the practical and scientific advantages of the previous \bicep/\keck\ experiments. Namely, 
\begin{itemize}
  \item the cryostat lends itself to complete assembly and disassembly by a small crew of experienced people; 
  \item the compact, on-axis optics allow for symmetric rotation of the telescope as a whole, an important method for probing systematics in the polarization-based measurement;
  \item the compact optics allow for ground-based characterization in the optical far-field;
  \item the compact receiver size allows installation of a comoving absorptive forebaffle that terminates beam sidelobes and reduces stray pickup;
% ZA-D: also mention compact comoving forebaffle to reduce stray pickup
  \item and the ability to cryogenically cool the optical elements and internal baffling minimizes the thermal photon noise seen by the detectors and leads to excellent stability.
\end{itemize}
% JW-D: Justin says this list is awkward. Are bullet points better?

This section provides a short overview of the \bicep3 instrument in its current, 2016 observing season state, offered here as both a recap before giving a status update in Sec.~\ref{sec:perf}, and to highlight features that will transfer over to a future \bicep3-class array as outlined in Sec.~\ref{sec:array}. 
A more detailed overview before the first (2015) season deployment is found in Ref.~\citenum{ahmed2014}, and an in-depth design and operation summary during the 2015 season is provided in Ref.~\citenum{thesis:wu}.

\subsection{Cryostat Receiver}

\begin{figure} [t]
  \begin{center}
    \begin{tabular}{c} %% tabular useful for creating an array of images 
      \includegraphics[width=.8\textwidth]{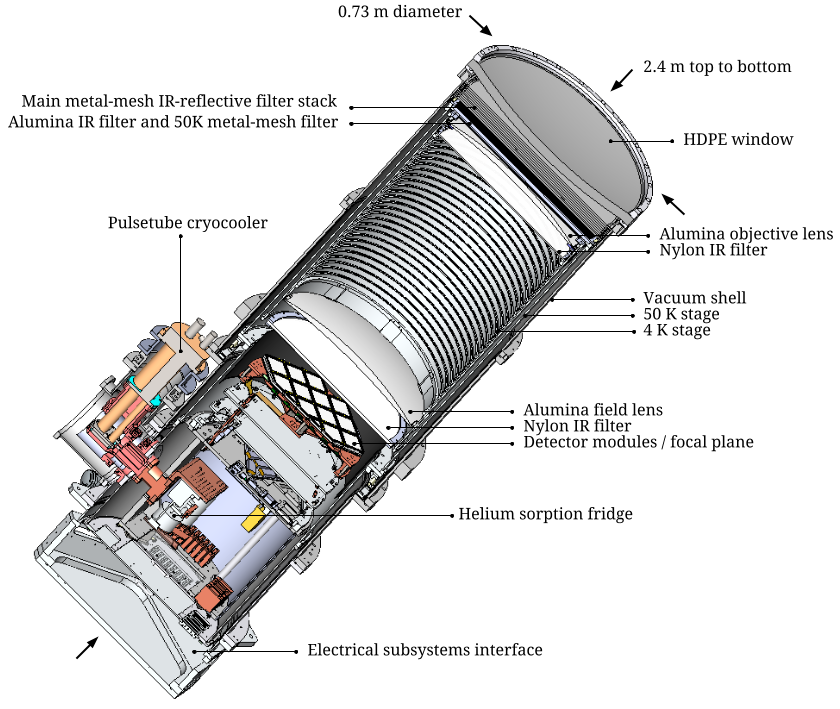}
    \end{tabular}
  \end{center}
  \caption[b3rxcutaway] 
  { \label{fig:b3_rx_cutaway} 
  Cutaway view of the \bicep3 cryostat receiver, with key elements labeled.
  }
\end{figure}

The \bicep3 receiver is a custom aluminum dry cryostat, \SI{2.4}{m} tall along the optical axis with a \SI{0.73}{m} outer diameter and totalling roughly \SI{550}{kg} without attached readout electronics and forebaffle. 
% IDB-D: weighing --> totalling
Fig.~\ref{fig:b3_rx_cutaway} shows a cutaway view of the \bicep3 CAD model in this state.
The top of the receiver is capped by a \SI{31.75}{mm} thick high-density polyethylene (HDPE) window through which light is collected, while the bottom end is used to interface to electronic subsystems. From the side protrudes a vacuum shell extension that houses the single PT415\footnote{Cryomech Inc., Syracuse, NY 13211, USA (\texttt{www.cryomech.com})} two-stage pulsetube cryocooler, which provides continuous cooling to $<$\SI{4}{K}.  
% KW-D: maybe drop the cooling to 3.3K - this number tells of the 4K loading and can’t be well interpreted without knowing the loading. Or can quote the load-free 4K-stage temperature

\begin{figure} [t]
  \begin{center}
    \begin{tabular}{c} %% tabular useful for creating an array of images 
      \includegraphics[width=.7\textwidth]{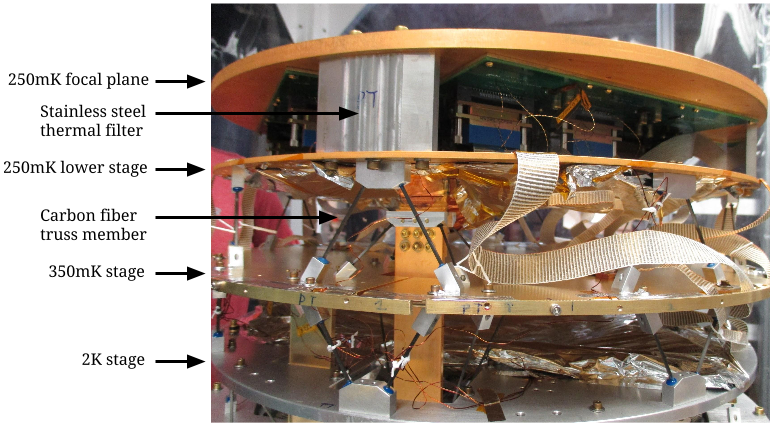}
    \end{tabular}
  \end{center}
  \caption[subk] 
  { \label{fig:subk} 
The \bicep3 sub-Kelvin stages exposed during assembly, with various components labeled.
  }
\end{figure}

The cryogenic design can be simply described as concentric cylindrical aluminum shells, each an increasingly cold and thermally isolating radiation shield that together protect the
sub-Kelvin detectors from warm radiation. 
% ZA-D: from warm radiation
% KLT-D: can remove this first sentence if needed. Anyone with cryo experience will understand.
Moving inward from the room-temperature vacuum shell are the nominal `\SI{50}{K}' and `\SI{4}{K}' stages, respectively cooled by the 1st and 2nd stages of the pulsetube. The pulsetube is thermally linked to the \SI{50}{K} and \SI{4}{K} stages in the main body of the cryostat via oxygen-free high-conductivity (OFHC) copper heat straps, including flexible foil and braid sections to minimize transmission of vibrational energy from the pulsetube. Within the \SI{4}{K} shield is a series of stacked planar `sub-Kelvin' stages (\SI{2}{K}, \SI{350}{mK}, \SI{250}{mK}) culminating in the \SI{250}{mK} focal plane (Fig.~\ref{fig:subk}). The various temperature stages are structurally supported and thermally isolated from each other by trusses of low thermal-conductivity materials (G10 fiberglass and carbon fiber). 

A three-stage ($^4$He/$^3$He/$^3$He) helium sorption fridge\footnote{Chase Research Cryogenics Ltd., Sheffield, S10~5DL, UK (\texttt{www.chasecryogenics.com})} is mounted on the \SI{4}{K} stage near the pulsetube and provides non-continuous cooling for all three sub-Kelvin stages. The three stages of the fridge contain \SI{55}{L} of $^4$He and \SI{33}{L}/\SI{4}{L} of $^3$He at standard temperature and pressure, able to cool \bicep3's detector modules and large copper focal plane from \SI{4}{K} to \SI{250}{mK} and then maintain operating temperatures for more than 48 hours
% ZA-D: two days --> 48 hours
before needing to re-cycle.

The \SI{250}{mK} stage is separated into two planar copper stages to provide thermal stability for the detectors. The lower level, closest to the \SI{350}{mK} stage, is actively temperature-controlled to \SI{269}{mK} by a resistive heater and acts as a buffer for the more isolated focal plane. Stainless steel blocks separate the two levels and provide low-pass thermal filtering. The focal plane itself is additionally temperature-controlled to \SI{274}{mK}.

\subsection{Detectors and Readout}

\bicep3's focal plane is populated by planar arrays of polarization-sensitive slotted antennas, coupled to transition edge sensor (TES) bolometers and centered at \SI{95}{GHz}, the same technology also fabricated by JPL/Caltech for \bicep2/\keck\,\arry\ and detailed in Ref.~\citenum{dets2015}. 
% KW-D: planar arrays of polarized, antenna-coupled transition edge sensor (TES) bolometers centered at 95 GHz, (bolometers are not polarized. perhaps slotted antennas sensitive to polarization?)
Each silicon detector tile contains an $8\times8$ array of pixels with each pixel made up of two co-located, orthogonally-polarized sub-antenna networks and two TES bolometers. 
% KW-D: clarify: each pixel has in it 8x8 pairs of co-located orthogonally laid-out sub-antennas, not just one pair/pixel
% KLT-D: pair differencing is confusing here, sounds like done in hardware. 
A Gaussian taper of each pixel's beam is achieved by tuning the summing tree that couples each bolometer to its sub-antenna network\cite{obrient2012}.
% KLT-D: didn't mention the taper summing
A major difference from \bicep2/\keck\,\arry\ is the installation method of detector tiles on the focal plane.
In \bicep2/\keck\ receivers, four detector tiles are directly wire bonded to the focal plane, and the supporting readout electronics sit adjacent at the focal plane's outer radius.
\bicep3\ instead packages each detector tile and its electronics in a more easily installed compact detector module. The concurrently presented Ref.~\citenum{spie:howard} describes these detector modules in more detail. 

%%% removed, defer to howards paper:
%%%
%The detector module houses the detector tile along with a quartz anti-reflective (AR) tile, a $\lambda/4$ spaced backshort, and circuit boards housing the 1st stage of multiplexing (MUX) superconducting quantum interference device (SQUID) readout chips, all in a magnetically-shielding niobium casing (Fig.~\ref{fig:module}). The bottom of each module is thermally connected only at the center to a copper baseplate, which is heatsunk to the main copper focal plane, in order to prevent formation of trapped magnetic flux as the module casing cools and superconducts. 
%% KW-D: in order to help expel internal magnetic fields as the module casing cools and superconducts. --> in order to prevent formation of trapped magnetic flux as the casing cools and superconducts.
%Each 79 x \SI{79}{mm} detector module therefore contains all of its required sub-Kelvin hardware, wire bonding, and SQUID readout chips in a footprint dominated by the detector tile itself and is quickly connected and heatsunk to the main focal plane with connectorized flex cables and mechanical fasteners.
%
%\begin{figure} [t]
%  \begin{center}
%    \begin{tabular}{c} %% tabular useful for creating an array of images 
%      \includegraphics[width=.7\textwidth]{img/module}
%    \end{tabular}
%  \end{center}
%  \caption[module] 
%  { \label{fig:module} 
%  Exploded view of one \bicep3 detector module. (need to update; make sure to specify up/down optical path orientation)
%  }
%\end{figure} 

Twenty such detector modules are tiled onto the \bicep3 focal plane, each containing 128 TES bolometers, for a total of 2560. Each detector tile contains 8 `dark' TES bolometers that are intentionally left uncoupled to antennas, 4 of which are read out as non-optical probes, giving a total of 2400 optically-coupled bolometers or, equivalently, 1200 polarization-sensitive detector pairs centered at \SI{95}{GHz}. This contrasts to a single \bicep2/\keck\,\arry\ receiver at \SI{150}{GHz}, which has four detector tiles for a total of 248 polarization-sensitive detector pairs. At \SI{95}{GHz}, a single \keck\,\arry\ receiver provides an even sharper comparison with 136 polarization-sensitive detector pairs, owing to its slower optics and smaller aperture than \bicep3.

\bicep3 continues to use time-domain multiplexed (TDM) readout of the TES bolometers, via SQUID MUX chips developed by NIST, in order to reduce the total number of lines that must enter the cryostat and their associated heat load. 
The signal current through each voltage-biased TES is inductively coupled to its own 1st-stage SQUID series array, forming one column of one multiplexed `row'. A pair of 11-row MUX chips is wire bonded in series in order to allow multiplexing through 22 rows. Each detector module contains 6 pairs of MUX chips, forming 6 columns of 22 rows, or 132 total 1st-stage channels. Each column's multiplexed output is inductively coupled to a separate SQUID series array on the \SI{4}{K} stage for additional amplification before exiting the cryostat.
\bicep3 uses MUX11d chips, a newer generation than those used by \bicep2/\keck\,\arry\cite{dekorte2003} that has a different row-switching mechanism. Details of the new design are found in the concurrently presented Ref.~\citenum{spie:howard}. 
Control of the MUX system and feedback-based readout of the TES data are via room temperature Multi-Channel Electronics (MCE) systems developed by the University of British Columbia\cite{mce2008}, again similar to those used in \bicep2/\keck\,\arry.

\subsection{Optics and IR Filtering}
\label{subsec:op}

Following the other \bicep/\keck\ experiments, \bicep3 targets degree-scale polarization features by using a compact, on-axis, two-refractor optical design. Its $f/1.7$ system with a \ang{27.4} field of view (FOV) is faster and wider than the $f/2.2$, FOV$\sim$\ang{15} optics of \bicep2/\keck\,\arry. It has a significantly larger aperture diameter of \SI{520}{mm}, vs. \SI{264}{mm} for \bicep2/\keck\,\arry, defined by a millimeter-wave absorbing aperture stop. Both refracting lenses and the aperture stop are heatsunk to the \SI{4}{K} stage, with the objective lens (warmest and furthest from the focal plane) operating at \SI{4.9}{K}. The lenses are made of 99.6\% pure alumina ceramic, which allows a thinner design and smaller thermal gradients than the HDPE plastic used for previous \bicep/\keck\ optics, crucial advantages for \bicep3's faster system and increased infrared (IR) thermal loading.

The much larger diameter of the \bicep3 receiver window ($\sim$\SI{670}{mm} clear), necessitated by the increase in aperture size, creates a large IR radiative heat load. IR power entering the cryostat is in excess of \SI{100}{W}, more than $2\times$ the cooling power of the 1st stage of the single PT415 pulsetube cooler. The first level of IR filtering therefore attempts to reject the majority of IR loading before it reaches the \SI{50}{K} stage, 
% KW-D: before it can be absorbed to the 50 K stage —> before it reaches the 50 K stage
and consists of a stack of 10 metal-mesh IR-reflective filters mounted just within the receiver window at room temperature. These filters are made up of $\mathcal{O}$(\SI{10}{$\mu$m}) capacitive squares created by laser ablation of aluminized thin plastic films (\SI{3.5}{$\mu$m} Mylar or \SI{6}{$\mu$m} polypropylene (PP)/polyethylene (PE) blend), as introduced in Ref.~\citenum{mesh2014}. In all, through both reflection and absorption/reemission, the metal-mesh filter stack allows $\sim$\SI{19}{W} to reach the \SI{50}{K} stage. One additional metal-mesh filter is mounted just behind the \SI{50}{K} alumina filter. The remaining power is absorbed by a series of three filters (and thereby removed by the pulsetube) in order to shield the sub-Kelvin detector stage: a \SI{10}{mm} thick alumina filter on the \SI{50}{K} stage, between the metal-mesh filter stack and the objective lens; a \SI{5}{mm} thick nylon filter on the \SI{4}{K} stage, between the two lenses; and a second \SI{4}{K} nylon filter, \SI{9.5}{mm} thick, between the field lens and the focal plane. These IR filters and lenses are labeled in Fig.~\ref{fig:b3_rx_cutaway}. 
% KLT-D: didn't mention ade filters
Finally, low-pass metal-mesh edge filters\cite{ade2006} with a \SI{4}{cm$^{-1}$} (\SI{120}{GHz}) cutoff are mounted onto each detector module to prevent any remaining out-of-band sub-millimeter blue leaks from coupling to the detector bolometers.

\section{Observing Site and Strategy}
\label{sec:obs}

The \bicep3 receiver is integrated with its telescope mount and various subsystems in the Dark Sector Laboratory (DSL) building at the South Pole, from where it observes the sky during the austral winter. 
% KLT-D: ``dark'' austral winter a little misleading
Support and logistics are managed through the US National Science Foundation's Amundsen-Scott South Pole Station, located \SI{1}{km} away, including satellite-based daily transfer of the full science data stream. This cold and remote observing site sits at an elevation of over \SI{2800}{m} on the Antarctic Plateau and provides extremely dry and stable atmospheric conditions, optimal for \bicep3's millimeter-wave observations.

\begin{figure} [t]
  \begin{center}
    \begin{tabular}{c} %% tabular useful for creating an array of images 
      \includegraphics[width=.5\textwidth]{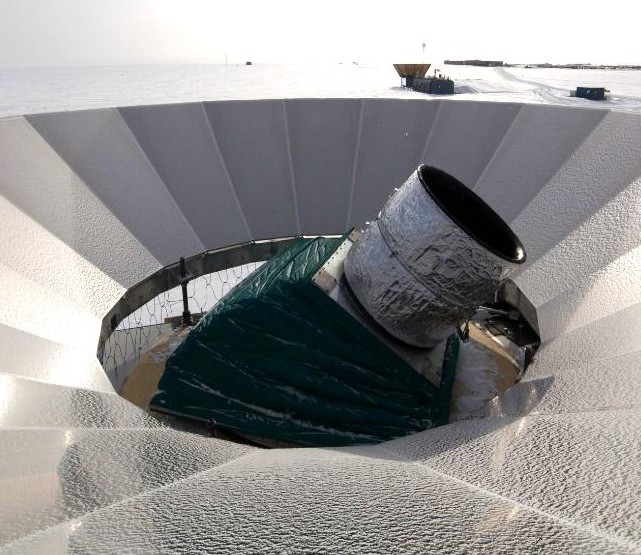}
    \end{tabular}
  \end{center}
  \caption[b3roof] 
  { \label{fig:b3_roof} 
  View of \bicep3 from the roof of DSL, showing the green insulating boot, comoving forebaffle, and reflective ground shield. The larger \keck\,\arry\ ground shield is visible in the background and the main Amundsen-Scott station extends along the horizon.
  }
\end{figure}

The warmed indoor environment of DSL is extended beyond the roof level by a flexible insulating boot so that only the receiver window is exposed to the outdoor environment (Fig.~\ref{fig:b3_roof}). 
%The majority of the telescope resides within the warmed indoor environment of DSL, which is extended beyond an opening in the roof by a flexible insulating boot such that only the receiver window is exposed to the outdoor environment (Fig.~\ref{fig:b3_roof}). 
A dry nitrogen gas system fills a protective membrane above the receiver window to prevent snow accumulation, and a cylindrical comoving microwave-absorptive forebaffle extends beyond the receiver window to intercept stray light. A reflective, stationary ground shield is fixed to the roof of DSL and acts as a second barrier against signal contamination from nearby ground sources.

\begin{figure} [t]
  \begin{center}
    \begin{tabular}{c} %% tabular useful for creating an array of images 
      \includegraphics[width=.6\textwidth]{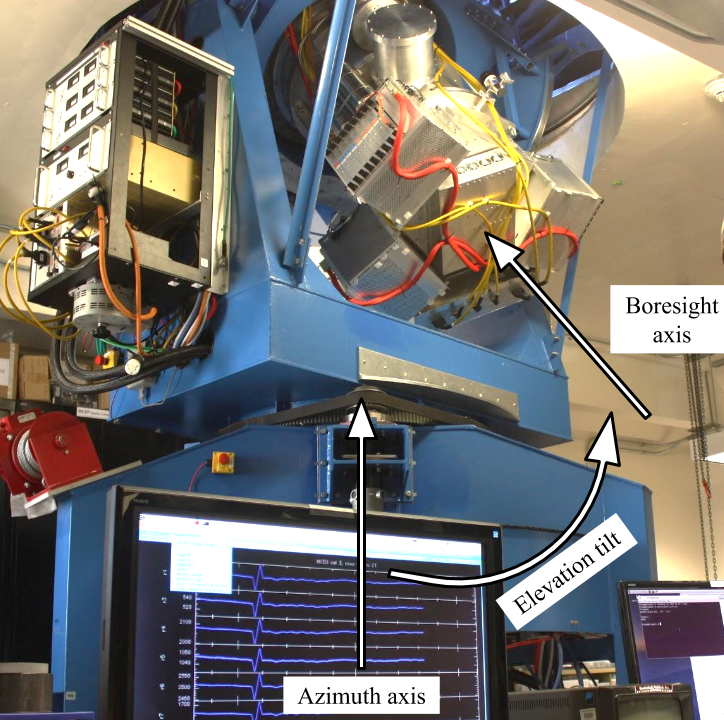}
    \end{tabular}
  \end{center}
  \caption[b3axes] 
  { \label{fig:b3_axes} 
  \bicep3 observing on the original \bicep\ mount, with the three axes of motion labeled.
  }
\end{figure}

\bicep3 continues to use the original three-axis \bicep1 mount in DSL that also supported \bicep2. By removing some non-essential structures, the mount was modified to fit \bicep3's larger size while retaining the same functions: ranges of \textgreater\ang{360} in azimuth, \ang{50}--\,\ang{90} (zenith) in elevation, and boresight rotation about the optical axis (Fig.~\ref{fig:b3_axes}). Due to the larger diameter of \bicep3 and the protruding vacuum shell extension that houses the pulsetube cooler, boresight rotation is limited by mechanical interference to a range of \ang{255}.
All electrical cabling, high-pressure helium gas hoses for the pulsetube, and other necessary lines are routed through cable carriers for the three axes of motion, with the exception of a 4-channel high-pressure rotary joint\footnote{Dynamic Sealing Technologies, Inc., Andover, MN 55304, USA (\texttt{www.dsti.com})} used to bypass the two helium hoses at the azimuth stage. By pressurizing two outer channels to act as buffers to atmosphere, the 4-channel rotary joint achieves much better leak performance than the 2-channel unit used during \bicep3's 2015 season; the optimal pulsetube system pressure is maintained by small, non-disruptive refills to the compressor less than once a month.

\bicep3 focuses its CMB observations on the same sky patch as the \keck\,\arry\ and previous \bicep\ experiments, spanning right ascension $\ang{-60}<\textrm{RA}<\ang{60}$ and declination $\ang{-70}<\delta<\ang{-40}$.
Its larger optical FOV results in an effective sky area of $\sim$ \SI{600}{deg$^2$}, larger than the $\sim$ \SI{400}{deg$^2$} area covered by \bicep2/\keck. 
% KW-D: I would change |α| < 60◦ to -60^{\circ} < alpha < 60^{\circ}
% ZA-D: Let’s drop the name “Southern Hole”. Just call it a 400 sq degree patch centered at RA xx, spanning yy in dec etc observed by BICEP/Keck, but larger by such and such 
The fundamental observing block is a constant-elevation `scanset' of 50 back-and-forth \SI{2.8}{deg\per s} fixed-center azimuth scans spanning \ang{64.4}. Each scanset is bookended by a \SI{26}{s} elevation nod of \ang{1.3} peak to peak for airmass-based detector gain calibration and $\sim$ \SI{1}{minute} of partial TES loadcurves (non-superconducting) to monitor detector parameters, for a total duration of $\sim$ 50 minutes. The azimuth center is adjusted every other scanset to track the changing RA of the sky patch, using that drift in combination with the field of view to fill in the full range in RA. Elevation is staggered by twenty \ang{0.25} offsets amongst the scansets to fill in coverage of the detector beam pattern. 

Observations are on a two-sidereal day cadence, including 40 CMB scansets and 7 scansets aimed at the galactic plane. Each two-day cycle uses one of four boresight angles within the allowed boresight envelope to help constrain polarization systematics: two sets of \ang{180} opposing orientations, offset from each other by \ang{45}, and clocked to optimize coverage symmetry of both the Stokes $Q$ and $U$ polarizations on the sky. The first \SI{6.5}{hrs} of each cycle is spent re-cycling the helium sorption fridge to keep the detectors cold for the rest of the two days.

\section{Current Performance}
\label{sec:perf}

\begin{figure} [t]
  \begin{center}
    \begin{tabular}{c} %% tabular useful for creating an array of images 
      \includegraphics[width=.82\textwidth]{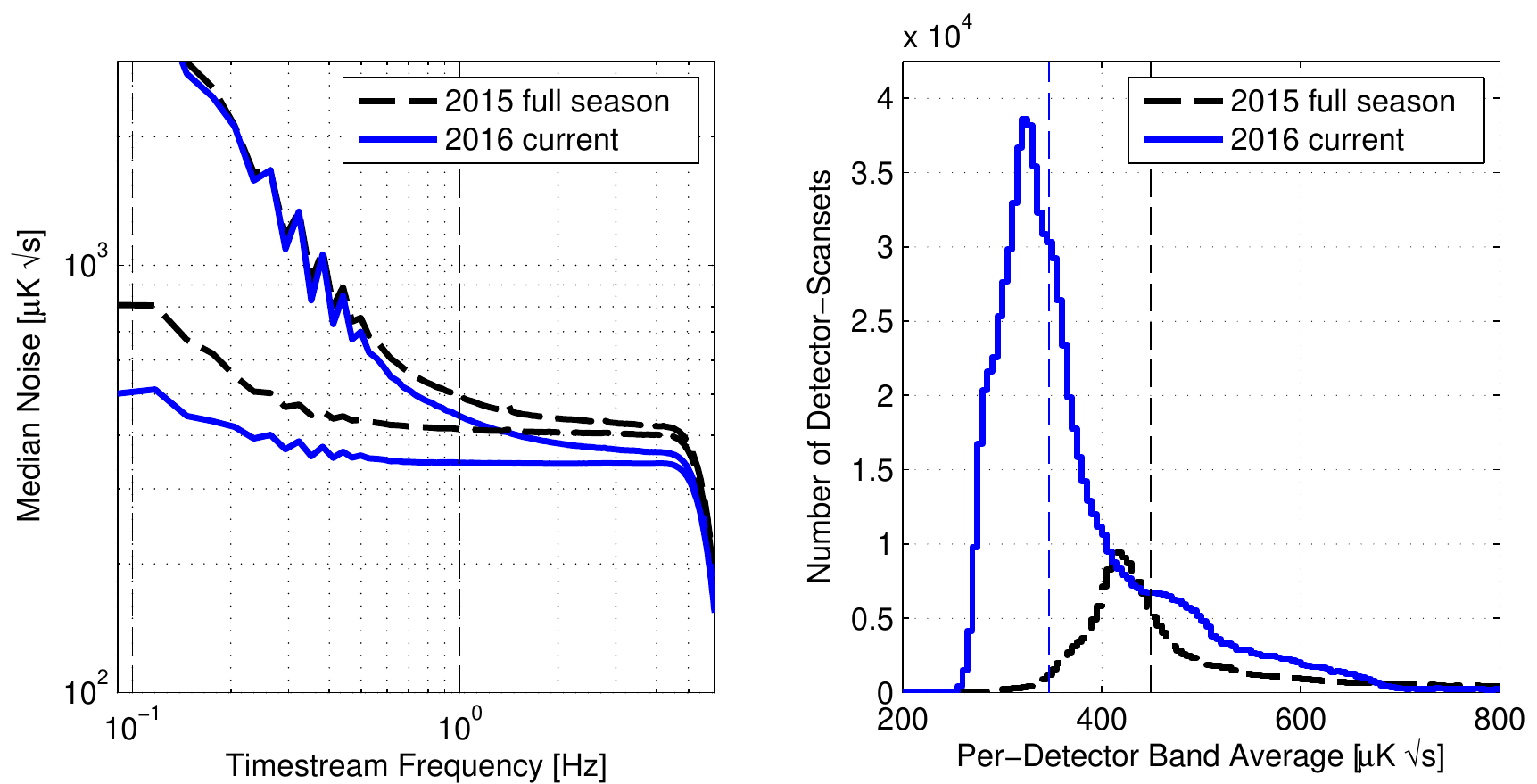}
    \end{tabular}
  \end{center}
  \caption[netspec] 
  { \label{fig:netspec} 
  \textit{Left}: Median per-detector noise spectra for \bicep3 2015 and 2016 season data, from both pair-summed and pair-differenced minimally processed timestreams (showing $1/f$ noise rejection of the differenced polarization measurement). 
  \textit{Right}: Histogram of the per-detector per-scanset noise, applying 3rd-order polynomial timestream filtering and averaged across the 0.1-\SI{1}{Hz} science band. Median values of \SI{449}{\ukrts} and \SI{347}{\ukrts} are marked by vertical dashed lines for the 2015 and 2016 data, respectively.
  Significantly increased detector population and observing duty cycle explain the much larger 2016 histogram amplitude, which will continue to grow as the season progresses.
  }
\end{figure}

\bicep3's first season performance, during the 2015 austral winter, is summarized in Ref.~\citenum{wu2015}. During that initial engineering season, 9 of the possible 20 detector modules were installed, for a total of 540 polarization-sensitive detector pairs centered at \SI{95}{GHz}. Yield after accounting for known hardware issues (e.g. suspected broken wire bonds, SQUID MUX chip yield) was 436 good polarization-sensitive pairs, or 81\%. 51 of the 104 lost pairs were attributed to a single anomalously low-yield detector module, potentially due to wire bond or other damage to the detector tile during installation and receiver cool-down. 
Median per-detector CMB noise-equivalent temperature (NET) for the 2015 season is estimated at \SI{449}{\ukrts}, based on polarization pair-differenced timestream data as shown in Fig.~\ref{fig:netspec} and explained later in this section.
% 81% is the RGL fp-data yield
%\ldots 67\% yield quoted in Kimmy's LTD paper? final overall cuts? erik said 38.4\% round2 pass frac
%\ldots mention 395 ukrts map-based NET from LTD paper? also agrees with Eric 2015 Fall meeting calc
% KLT-D: make clear that the NET reported is for FIRST season

\bicep3 underwent a number of improvements during the 2015/16 austral summer in preparation for its second season in the 2016 austral winter, which is the first dedicated science season. As originally planned, the focal plane was populated with the full complement of 20 detector modules, including 12 new detector modules and the removal of the anomalously low-yield module mentioned above.
%(Fig.~\ref{fig:twenty})
 Additionally, several important changes were made in response to data and operations knowledge collected during the 2015 season. 

%%% deferd to howard's:
%\begin{figure} [t]
%  \begin{center}
%    \begin{tabular}{c} %% tabular useful for creating an array of images 
%      \includegraphics[width=.5\textwidth]{img/twenty}
%    \end{tabular}
%  \end{center}
%  \caption[twenty] 
%  { \label{fig:twenty} 
%The \bicep3 focal plane during assembly, showing all 20 detector modules.
%  }
%\end{figure} 
 
First, cryogenic improvements have added over 20\% to the helium sorption fridge duty cycle and allow the current two-day observing schedule cadence. During the 2015 season, continuous observing time was limited by the fridge to \SI{12}{hrs}. This cryogenic fix came in two parts, both vetted by parallel lab testing in North America during \bicep3's 2015 season observations. Lowering of the helium sorption fridge bath temperature by \SI{0.6}{K}, to \SI{3.5}{K}, was achieved by increasing thermal conduction through the copper heatstrap to the pulsetube's 2nd stage and thereby allowed increased helium condensation during the fridge re-cycling. Furthermore, a foil shielding included on a small number of electrical cables was measured to give an excess \SI{200}{$\mu$W} parasitic conductive heat load to the nominal \SI{350}{mK} stage, at least tripling the total expected load. Removal of this incorrectly specified shielding resulted in a large \SI{280}{mK} reduction on that stage, allowing it to reach \SI{360}{mK}, which in turn allowed lowering the focal plane operating temperature from \SI{313}{mK} to \SI{274}{mK} for the 2016 season.
% 200pW number from ethan

Second, two faulty metal-mesh IR-reflective filters were replaced for 2016 and have significantly reduced broad and polarization-dependent scattering of the 95 GHz signal. This has been measured both as a per-detector 26\% median reduction of loading from the comoving forebaffle, and as an improvement in detector loading median mismatch from $\sim$~10\% to $<3$\% between co-located orthogonally polarized detectors when observing broad sources (forebaffle and atmosphere). Lab inspection has shown these faulty filters to be plagued by incomplete and inconsistent laser ablation of the aluminization on micron scales, now easily remedied with replacements and improved screening.
% see slides/posting links from 2016 May Caltech collab meeting

Finally, 2015 season data showed \bicep3 to be sensitive to \SI{450}{MHz} interference from the main South Pole Station radio system, owing to its larger receiver aperture than \bicep2/\keck\,\arry\ (Sec.~\ref{subsec:op}) and correspondingly lower high-pass waveguide cutoff frequency. The interference caused a mostly ground-fixed signal, but also contributed to transients in the feedback-based detector readout. For the 2016 season, internal RF shielding improvements at the focal plane have given a $\sim$~\SI{10}{dB} reduction in RF signal sensitivity, and changes at the main station (reduced output power and a directional antenna) have reduced the signal strength at DSL by $\sim$~\SI{35}{dB}, for a total $\sim$~\SI{45}{dB} reduction.
% see slides/posting links from 2016 May Caltech collab meeting

As is routine for the \bicep/\keck\ experiments, \bicep3 was subject to several calibration measurements during the 2015/16 austral summer in order to characterize the instrument before the 2016 observation season. 
Per-detector spectra were measured for all 20 modules with a custom Martin-Puplett Fourier Transform Spectrometer (FTS)\cite{karkare2014} mounted to the window of \bicep3 while in the mount.
%%% Defer to howard's:
%%%
% The averaged spectrum across all detectors is shown in Fig.~\ref{fig:fts_avg}.
% FTS posting from Sinan:
% http://bicep.rc.fas.harvard.edu/bicep3/analysis_logbook/20160405_B3_SP_FTS/
%
Optical efficiency through the receiver was measured with an aperture-filling liquid nitrogen load mounted above the receiver window, which also acted as an ambient temperature source when unfilled. 
Results from both measurements can be found in the concurrently presented Ref.~\citenum{spie:howard}.
%%% Defer to howard's:
%%%
%The average efficiency across all detectors is $\sim$24\%, assuming the average bandwidth. 
% dP/dT from Sinan posting:
% http://bicep.rc.fas.harvard.edu/bicep3/analysis_logbook/20160105_B3_SP_run08_OE/
% used 26.3 GHz BW to calculate OE percentage
% Using per-det actual FTs spectra normalized to 1, get more like 43% (!)
Far-field beam characteristics were measured over several weeks using a chopped thermal source mounted high above the nearby Martin A. Pomerantz Observatory building (the location of \keck\,\arry) and are covered in detail in the concurrently presented Ref.~\citenum{spie:kirit}.

%%% Defer to howard's:
%%%
%\begin{figure} [t]
%  \begin{center}
%    \begin{tabular}{c} %% tabular useful for creating an array of images 
%      \includegraphics[width=.38\textwidth]{img/fts_avg}
%    \end{tabular}
%  \end{center}
%  \caption[ftsavg] 
%  { \label{fig:fts_avg} 
%The averaged response spectrum across all \bicep3 detectors, with a band center of \SI{93.3}{GHz} and a bandwidth of \SI{26.3}{GHz}.
%  }
%\end{figure} 

As of writing in early June 2016, \bicep3 has observed for $\sim2.5$ months in its second season and accumulated roughly 1224 receiver-hours of new CMB data. 
% 1692 tags between march 15 and early June used in NET hist. multiply by 43.4 minutes t_onsrc.
Overall detector yield due to known hardware issues is 951 good polarization-sensitive pairs, or 79\%. 180 of the 249 lost pairs are due to electrical open circuits that developed on 7 SQUID MUX row-select lines shared by a group of 5 detector modules. The cause of the bad connections is yet to be determined and only became present after cryogenic cooling of the receiver. Because these row-select lines are run in series, it is possible that poor wire bond adhesion within a single detector module could be the culprit. 
The number of lost detector pairs is magnified due to polarization pairing across adjacent MUX rows (pairs are wired along the same MUX column, rather than row, sharing common parameters such as SQUID and TES biases). 
% this is fp_data RGL yield

The median noise spectrum and per-detector, per-scanset noise distribution is shown in Fig.~\ref{fig:netspec}, based on the timestreams for all CMB data so far during the 2016 season. The timestream data has been differenced between polarization pairs, with basic cuts applied so that only good pairs remain (based on both the known hardware yield and preliminary data-quality cuts). Pair-summed spectra are also shown to demonstrate the $1/f$ noise rejection of the pair-difference polarization data. Conversion to CMB temperature is done by calibration of a \bicep3 CMB temperature map to \SI{100}{GHz} Planck data. 
The histogram shows science-band average noise after filtering the timestreams with a 3rd-order polynomial during every unidirectional azimuth scanning movement (100 moves per scanset) to mimic final map making.
\bicep3's 2015 season data is also shown for comparison. 

An improvement in per-detector sensitivity is clear, summarized by the improved per-detector median band-average NET of \SI{347}{\ukrts} for the 2016 season versus \SI{449}{\ukrts} from the 2015 season data (right panel of Fig.~\ref{fig:netspec}). However, \keck\,\arry\ \SI{95}{GHz} data shows a median \SI{280}{\ukrts} with the same treatment. Reduction of scattering and the resulting detector photon load from warm surfaces, along with the reduction of RF interference, both detailed earlier in this section, can explain \bicep3's improvement in the 2016 season, but a detailed quantification and an understanding of the performance relative to \keck\,\arry\ detectors is ongoing.
Combining the band-average NET values across all detectors for each scanset gives a \bicep3 2016 median array NET of \SI{9.91}{\ukrts}. This value accounts for hardware yield (79\%) and for preliminary data-quality cuts (currently a $\sim$45\% loss). \bicep3 2015 data gives a much higher median array NET of \SI{24.6}{\ukrts} due to the higher per-detector NET, the only partially populated focal plane, and a higher ($\sim$58\%) loss to data-quality cuts.
% http://bicep.rc.fas.harvard.edu/bkcmb/analysis_logbook/analysis/20160614_b3_nethist_pt2/

All above quoted NET values in this section are based on timestream data. This is an important distinction and the values are not meant to be directly comparable to map-based NET estimates. 
Specifically, Ref.~\citenum{wu2015} and Ref.~\citenum{spie:howard} quote map-based \bicep3 per-detector NET estimates of \SI{395}{\ukrts} (2015 season) and \SI{333}{\ukrts} (2016 season). 
The difference arises from details of the two NET estimations. 
The map-based value is an estimate based on the standard deviation across all map pixels of a polarization noise map, accounting for the weighted integration time of each pixel, and is therefore insensitive to details of individual detector spectra and assumes a Gaussian noise distribution that will not necessarily reproduce a median of the distributions shown in Fig.~\ref{fig:netspec}. 
Additionally, the timestream-based value averages across 0.1--\SI{1}{Hz}, not the full range of scales considered in the map-based method described.
The two methods converge more closely for the total \bicep3 array NET, for which the map-based method gives \SI{9.72}{\ukrts}\cite{spie:howard}.  

\section{Future Multi-Receiver Array}
\label{sec:array}

Plans have begun to upgrade the \keck\,\arry\ to a 3rd-generation, or ``Stage 3'', experiment with \bicep3-class receivers. The \keck\,\arry, the culmination of the \bicep2/\keck\ 2nd-generation, demonstrates the scalability of 
% ZA-D: proven is sort of a proposal word. You want to use objective adjectives in a paper.
\bicep2-class receiver technology and the efficiency of frequency expansion through an array model in which the optics and detectors of each receiver can be tuned to a single frequency band. For the 2014 observing season, two of the five \SI{150}{GHz} \keck\,\arry\ receivers were replaced with \SI{95}{GHz} detectors and matching optics. That additional data allowed a more extensive model of the polarized galactic foreground, and for the first time showed B-mode constraints to be the most powerful on the IGW signal (beyond constraints derived from CMB temperature and other evidence)\cite{BKVI}. 
% KW-D: and for the first time showed B-mode constraints to be the most powerful on the IGW signal. (This sentence is hard to understand. perhaps: for the first time yielded a more powerful constraint on IGW with B-mode measurements.)
In the following 2015 observing season, two more \SI{150}{GHz} \keck\,\arry\ receivers were refitted, this time with \SI{220}{GHz} detectors and optics, further broadening the spectral coverage of the combined data set by reaching to the higher-frequency polarized dust signal. 
% and already showing excellent signal-to-noise on the polarized dust signal that the higher frequency is meant to probe.
% ZA-D: is there a keck proceeding/talk quoting 220 map depth? need ref.
Most recently, now that \bicep3 is fully populated at \SI{95}{GHz} for the current 2016 observing season, the two \SI{95}{GHz} \keck\,\arry\ receivers were also refitted to \SI{220}{GHz}, for a total of four at \SI{220}{GHz}.

%\bicep3 is proving to be a successful evolution 
% ZA-D: this is too qualitative. skip proving to be a succesful evolution. Just 10x detector count makes it an ideal building block for expansion 
With its 5$\times$ higher per-receiver detector count, the \bicep3 design acts as a building block for this 3rd-generation multi-frequency CMB polarimeter that we have chosen to name \bicep\,Array. In order to continue to further constrain the IGW signal and aim for primordial B-mode detection, \bicep\,Array will focus its increased total detector count ($\sim$~12$\times$ greater than \keck\,\arry)
% 12X from 5*512=Keck, BA=table in proposal
on the same sky patch from the South Pole. This shared sky patch with all other \bicep/\keck\ experiments covers $\sim$1\% of the sky, sufficient area to search for the degree-scale (multipole $l=80$) recombination bump B-mode signal. With the large amount of accumulated observations, it remains the preferred patch 
% ZA-D: not prefered patch because it’s 1%. prefered patch because we have lots of 150 and now 220 data on it.
to continue with deeper, multi-frequency observations for targeted degree-scale IGW signal detection and improved galactic foreground separation.

\bicep\,Array will be 4 new \bicep3-class receivers with frequency band centers spanning 35 to \SI{270}{GHz}. The design of a \bicep3-class receiver was detailed in Sec.~\ref{sec:inst}, showing how \bicep3 expanded to higher detector count and larger aperture while maintaining the advantages of the compact, on-axis, cold refractor design shared by the \bicep/\keck\ program. A number of changes are expected to improve on the \bicep3 design. Testing of the metal-mesh IR-reflective filter design has been ongoing since its first development for \bicep3, and potential changes include double-sided aluminization and laser patterning so that fewer filters are required to achieve the same IR rejection. Heatsinking the metal-mesh filter stack to \SI{50}{K}, rather than the room temperature vacuum shell as in \bicep3, could also reduce IR loading from the emissive substrate films. 

\begin{figure} [t]
  \begin{center}
    \begin{tabular}{c} %% tabular useful for creating an array of images 
      \includegraphics[width=1\textwidth]{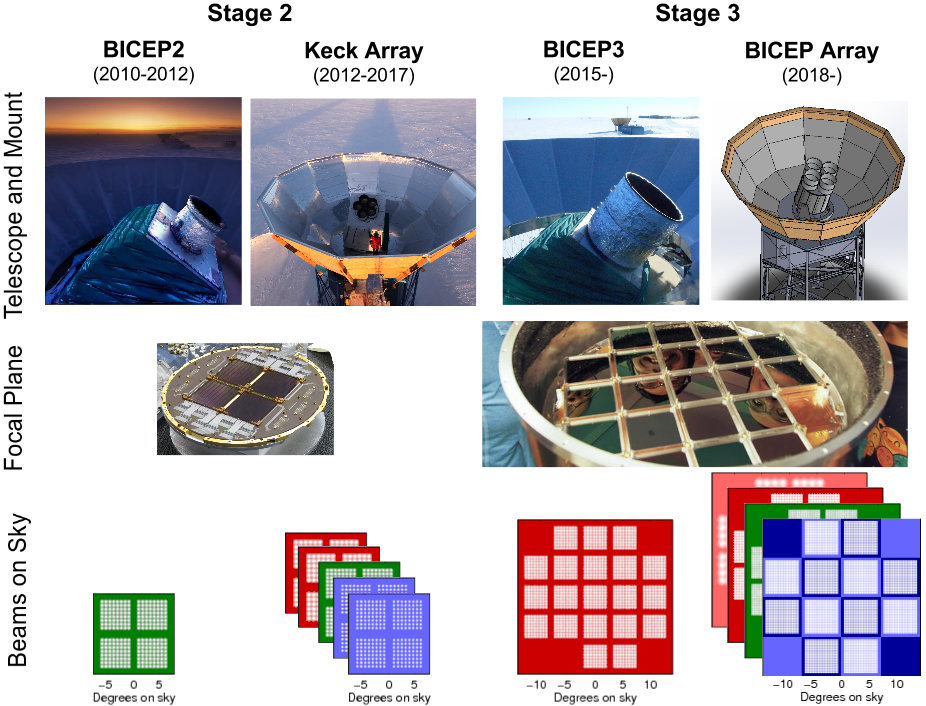}
    \end{tabular}
  \end{center}
  \caption[bkhistory] 
  { \label{fig:bkhistory} 
  Summary of the 2nd and 3rd-generations of the \bicep/\keck\ program. The bottom row shows the beam patterns on the sky with a common scale. 
  }
\end{figure}

Some changes are also expected for the detector modules used in \bicep\,Array. The four receivers will separately have focal planes centered on 35, 95, and \SI{150}{GHz}, and lastly a 220/\SI{270}{GHz} hybrid. Each focal plane will be made up of \bicep3-style detector module packages but will use 6'' rather than 4'' silicon wafer processes and require many fewer than 20 modules to fill the focal plane, thereby reclaiming focal plane area lost to the module borders. 
Table~\ref{tab:numbers} shows the nominal number of detectors that will fit into each focal plane of \bicep\,Array.
The 220/\SI{270}{GHz} dust-control receiver will have detectors at the two frequencies in a checkerboard array on the focal plane, based on the current \SI{220}{GHz} \keck\,\arry\ detectors and a lab-tested \SI{270}{GHz} design that will be introduced into \keck\,\arry\ for the 2017 season. 
The extremely high 220/\SI{270}{GHz} detector count motivates frequency domain multiplexing readout schemes as a possible alternative to \bicep3's SQUID-based TDM readout (which will continue to be used for the lower frequency receivers). 
The \SI{35}{GHz} receiver will help constrain galactic synchrotron emission at the level required for \bicep\,Array's sensitivity; prototyping and testing of the \SI{35}{GHz} detectors is currently underway. 
Fig.~\ref{fig:bkhistory} shows a comparison of the proposed \bicep\,Array design to its predecessors, including the focal plane size and frequency mix. 
% ZA-D: This should all be in a table (detector counts). Suggest using Jamie Bock’s table.

\begin{table}
  \centering
  \begin{tabular}{l c c c}
    \toprule
    Experiment & Frequency Band (GHz) & TES Detector Count & Survey Weight Per Year  \\
    & & Per Receiver &  Per Receiver ($\mu$K$^{-2}$\,yr$^{-1}$) \\
    \toprule
    \keck\,\arry  & \p{0}95 & \p{0}\tbf{288} & \p{0}\tbf{24000} \\
                  &     150 & \p{0}\tbf{512} & \p{0}\tbf{30000} \\
		  &     220 & \p{0}\tbf{512} & \p{00}\tbf{2000} \\
		  &     270 &      \p{0}512  &      \p{000}800  \\
    \midrule
    \bicep3       & \p{0}95 &     \tbf{2560} &     \p{*}213000\footnotemark\\
    \midrule
    \bicep\,Array & \p{0}35 &      \p{0}384  &      \p{0}37000  \\
                  & \p{0}95 &          3456  &          287000  \\
                  &     150 &          7776  &          453000  \\
		  &     220 &          9408  &      \p{0}37000  \\
		  &     270 &          9408  &      \p{0}15000  \\
    \bottomrule
  \end{tabular}
  \caption{Nominal number of TES detectors and survey weight per year for each of the four \bicep\,Array receivers (\SI{220}{GHz} and \SI{270}{GHz} make up one receiver), and a comparison to \bicep3 and \keck\,\arry. 
    Bolded detector counts are from fielded receivers.
    Maximum \keck\,\arry\ (\bicep\,Array) population across 5 (4) receivers is 2560 (30432) detectors. 
    Bolded survey weights are achieved results, indicating performance taking into account all real world observing inefficiencies (see text for details). Other survey weights are scaled from the bolded values.
    }
  \label{tab:numbers}
\end{table}

\bicep\,Array will use a new mount at the South Pole to hold all 4 \bicep3-class receivers, each arranged around a common boresight, as the increased receiver size cannot be accommodated by the existing \keck\,\arry\ mount (originally built for the D\textsc{asi} experiment). The use of high-pressure helium rotary joints, operating successfully with \bicep3, will help reduce the total size of the new mount by avoiding a significant volume of helium hose in the azimuth and theta stages. \bicep\,Array will replace the \keck\,\arry\ and observe simultaneously with the ongoing \bicep3 program. The \SI{35}{GHz} receiver is under development and is planned to be the first of the \bicep\,Array, beginning as early as the 2018 observing season. Three \keck\,\arry\ receivers outfitted at \SI{270}{GHz} will fill the remaining spots of the new mount. Before the second observing season, the 95 and \SI{150}{GHz} receivers will be installed, and finally the 220/\SI{270}{GHz} receiver will complete the array for the third and subsequent observing seasons. 

A useful metric to quantify the CMB polarization sensitivity and the statistical sensitivity to $r$ that will be achieved with \bicep\,Array is the survey weight $W=2A/N^2$ (units of $\mu$K$^{-2}$), where $A$ is the map area and $N$ is the polarization map depth (e.g. units of $\mu$K$\cdot$degrees). 
Conveniently, survey weight scales linearly with detector count and integration time, and Table~\ref{tab:numbers} shows a comparison of survey weights per year for \keck\,\arry\ and \bicep\,Array receivers. 
The \keck\,\arry\ \SI{95}{GHz} survey weight is based on the 2014 season published result\cite{BKVI}, 
% BK6 quoted 47000 survey weight for two receiver years --> 23500 --> 24000/rx
while the \SI{150}{GHz} survey weight is based on both \bicep2 performance\cite{BKI} and the latest published 3 receiver-years of \keck\,\arry\cite{BKVI}.
% BK1 gives ~33000/rx for BICEP2
% BK6 gives ~23000/rx for Keck when subtracted by BKP
The \keck\,\arry\ \SI{220}{GHz} survey weight is based on preliminary results from the first season (2015) of receivers at that frequency. 
% 20151021_keck2015_pager gives ~1540/rx for dk jack and others
% 20160425_bk15pager gives ~1750/rx for no jack
These achieved values are then simply scaled by the squared ratio of nominal per-detector NET values (taking into account e.g. detector parameters, loading from optical elements and the atmosphere) to 35 and \SI{270}{GHz}, and the \bicep\,Array values are further scaled by the ratio of detector counts.
In this way, all values take into account all real world observing inefficiencies of the \bicep2/\keck\ experiments. 
From this treatment it is clear that the \bicep\,Array represents an order of magnitude upgrade in total raw experimental sensitivity versus \keck\,\arry.

\footnotetext{
      The quoted \bicep3 survey weight is scaled from the \keck\,\arry\ \SI{95}{GHz} value, reflecting the eventual expectation for \bicep3. Adjusting by the \bicep3 and \keck\,\arry\ median per-detector band average NET values quoted in Sec.~\ref{sec:perf} gives an estimated \bicep3 survey weight of \SI{139000}{$\mu$K$^{-2}$} for 2016. 
    }

%% Remove reference to sigma(r) estimates:
% scaling achieved per-detector \bicep2/\keck\ sensitivities to the proposed detector counts, and accounting for galactic foregrounds and the lensing signal, \bicep\,Array is projected to reach $\sigma(r)<0.005$ after four seasons. For comparison, the latest combined \bicep/\keck\ results give $\sigma(r)=0.024$\cite{BKVI}.

\section{Conclusion}

These proceedings have summarized the instrument design and observation strategy of the \bicep3 polarimeter, and reported on changes and current performance for its ongoing second season. 
Important changes included full detector population of the focal plane, cryogenic improvements that led to increased observation duty cycle, replacement of under-performing IR filters, and a significant reduction in RF interference.
Based on the first $\sim1224$~hours of new CMB data this season (2016), \bicep3 has a timestream-based median per-detector NET of \SI{347}{\ukrts} and an array NET of \SI{9.91}{\ukrts} for the whole instrument, an improvement over the 2015 per-detector median of \SI{449}{\ukrts}. 

Plans were also outlined for an upgrade to the multi-frequency \keck\,\arry, named \bicep\,Array. 
\bicep\,Array will be four \bicep3-class receivers spanning 35 to \SI{270}{GHz} and begin a staggered deployment as soon as 2018. 
It will use a new mount at the existing South Pole \keck\,\arry\ site, and have more than an order of magnitude greater survey weight per receiver versus \keck\,\arry.

%% file: main.bbl
\begin{thebibliography}{10}

\bibitem{pbear2014}
{The P\textsc{olarbear} Collaboration}, Ade, P. A.~R., Akiba, Y., Anthony,
  A.~E., Arnold, K., Atlas, M., Barron, D., Boettger, D., Borrill, J., Chapman,
  S., Chinone, Y., Dobbs, M., Elleflot, T., Errard, J., Fabbian, G., Feng, C.,
  Flanigan, D., Gilbert, A., Grainger, W., Halverson, N.~W., Hasegawa, M.,
  Hattori, K., Hazumi, M., Holzapfel, W.~L., Hori, Y., Howard, J., Hyland, P.,
  Inoue, Y., Jaehnig, G.~C., Jaffe, A.~H., Keating, B., Kermish, Z., Keskitalo,
  R., Kisner, T., Jeune, M.~L., Lee, A.~T., Leitch, E.~M., Linder, E., Lungu,
  M., Matsuda, F., Matsumura, T., Meng, X., Miller, N.~J., Morii, H., Moyerman,
  S., Myers, M.~J., Navaroli, M., Nishino, H., Orlando, A., Paar, H., Peloton,
  J., Poletti, D., Quealy, E., Rebeiz, G., Reichardt, C.~L., Richards, P.~L.,
  Ross, C., Schanning, I., Schenck, D.~E., Sherwin, B.~D., Shimizu, A.,
  Shimmin, C., Shimon, M., Siritanasak, P., Smecher, G., Spieler, H., Stebor,
  N., Steinbach, B., Stompor, R., Suzuki, A., Takakura, S., Tomaru, T., Wilson,
  B., Yadav, A., and Zahn, O., ``{A Measurement of the Cosmic Microwave
  Background B-mode Polarization Power Spectrum at Sub-degree Scales with
  POLARBEAR},'' {\em The Astrophysical Journal}~{\bf 794},  171 (October 2014).

\bibitem{sptpol2015}
Keisler, R., Hoover, S., Harrington, N., Henning, J.~W., Ade, P. A.~R., Aird,
  K.~A., Austermann, J.~E., Beall, J.~A., Bender, A.~N., Benson, B.~A., Bleem,
  L.~E., Carlstrom, J.~E., Chang, C.~L., Chiang, H.~C., Cho, H.-M., Citron, R.,
  Crawford, T.~M., Crites, A.~T., de~Haan, T., Dobbs, M.~A., Everett, W.,
  Gallicchio, J., Gao, J., George, E.~M., Gilbert, A., Halverson, N.~W.,
  Hanson, D., Hilton, G.~C., Holder, G.~P., Holzapfel, W.~L., Hou, Z., Hrubes,
  J.~D., Huang, N., Hubmayr, J., Irwin, K.~D., Knox, L., Lee, A.~T., Leitch,
  E.~M., Li, D., Luong-Van, D., Marrone, D.~P., McMahon, J.~J., Mehl, J.,
  Meyer, S.~S., Mocanu, L., Natoli, T., Nibarger, J.~P., Novosad, V., Padin,
  S., Pryke, C., Reichardt, C.~L., Ruhl, J.~E., Saliwanchik, B.~R., Sayre,
  J.~T., Schaffer, K.~K., Shirokoff, E., Smecher, G., Stark, A.~A., Story,
  K.~T., Tucker, C., Vanderlinde, K., Vieira, J.~D., Wang, G., Whitehorn, N.,
  Yefremenko, V., and Zahn, O., ``{Measurements of Sub-degree B-mode
  Polarization in the Cosmic Microwave Background from 100 Square Degrees of
  SPTpol Data},'' {\em The Astrophysical Journal}~{\bf 807},  151 (July 2015).

\bibitem{BKI}
{BICEP2 Collaboration}, Ade, P. A.~R., Aikin, R.~W., Barkats, D., Benton,
  S.~J., Bischoff, C.~A., Bock, J.~J., Brevik, J.~A., Buder, I., Bullock, E.,
  Dowell, C.~D., Duband, L., Filippini, J.~P., Fliescher, S., Golwala, S.~R.,
  Halpern, M., Hasselfield, M., Hildebrandt, S.~R., Hilton, G.~C., Hristov,
  V.~V., Irwin, K.~D., Karkare, K.~S., Kaufman, J.~P., Keating, B.~G.,
  Kernasovskiy, S.~A., Kovac, J.~M., Kuo, C.~L., Leitch, E.~M., Lueker, M.,
  Mason, P., Netterfield, C.~B., Nguyen, H.~T., O'Brient, R., Ogburn, R.~W.,
  Orlando, A., Pryke, C., Reintsema, C.~D., Richter, S., Schwarz, R., Sheehy,
  C.~D., Staniszewski, Z.~K., Sudiwala, R.~V., Teply, G.~P., Tolan, J.~E.,
  Turner, A.~D., Vieregg, A.~G., Wong, C.~L., and Yoon, K.~W., ``{Detection of
  $B$-Mode Polarization at Degree Angular Scales by BICEP2},'' {\em Phys. Rev.
  Lett.}~{\bf 112},  241101 (June 2014).

\bibitem{BKVI}
{Keck Array and BICEP2 Collaborations}, Ade, P. A.~R., Ahmed, Z., Aikin, R.~W.,
  Alexander, K.~D., Barkats, D., Benton, S.~J., Bischoff, C.~A., Bock, J.~J.,
  Bowens-Rubin, R., Brevik, J.~A., Buder, I., Bullock, E., Buza, V., Connors,
  J., Crill, B.~P., Duband, L., Dvorkin, C., Filippini, J.~P., Fliescher, S.,
  Grayson, J., Halpern, M., Harrison, S., Hilton, G.~C., Hui, H., Irwin, K.~D.,
  Karkare, K.~S., Karpel, E., Kaufman, J.~P., Keating, B.~G., Kefeli, S.,
  Kernasovskiy, S.~A., Kovac, J.~M., Kuo, C.~L., Leitch, E.~M., Lueker, M.,
  Megerian, K.~G., Netterfield, C.~B., Nguyen, H.~T., O'Brient, R., Ogburn,
  R.~W., Orlando, A., Pryke, C., Richter, S., Schwarz, R., Sheehy, C.~D.,
  Staniszewski, Z.~K., Steinbach, B., Sudiwala, R.~V., Teply, G.~P., Thompson,
  K.~L., Tolan, J.~E., Tucker, C., Turner, A.~D., Vieregg, A.~G., Weber, A.~C.,
  Wiebe, D.~V., Willmert, J., Wong, C.~L., Wu, W. L.~K., and Yoon, K.~W.,
  ``{Improved Constraints on Cosmology and Foregrounds from BICEP2 and Keck
  Array Cosmic Microwave Background Data with Inclusion of 95 GHz Band},'' {\em
  Phys. Rev. Lett.}~{\bf 116},  031302 (January 2016).

\bibitem{kam2016}
Kamionkowski, M. and Kovetz, E.~D., ``{The Quest for B Modes from Inflationary
  Gravitational Waves},'' {\em Ann. Rev. Astron. Astrophys.}~{\bf 54}
  (September 2016).

\bibitem{barkats2014}
{Barkats}, D., {Aikin}, R., {Bischoff}, C., {Buder}, I., {Kaufman}, J.~P.,
  {Keating}, B.~G., {Kovac}, J.~M., {Su}, M., {Ade}, P.~A.~R., {Battle}, J.~O.,
  {Bierman}, E.~M., {Bock}, J.~J., {Chiang}, H.~C., {Dowell}, C.~D., {Duband},
  L., {Filippini}, J., {Hivon}, E.~F., {Holzapfel}, W.~L., {Hristov}, V.~V.,
  {Jones}, W.~C., {Kuo}, C.~L., {Leitch}, E.~M., {Mason}, P.~V., {Matsumura},
  T., {Nguyen}, H.~T., {Ponthieu}, N., {Pryke}, C., {Richter}, S., {Rocha}, G.,
  {Sheehy}, C., {Kernasovskiy}, S.~S., {Takahashi}, Y.~D., {Tolan}, J.~E., and
  {Yoon}, K.~W., ``{Degree-scale Cosmic Microwave Background Polarization
  Measurements from Three Years of BICEP1 Data},'' {\em The Astrophysical
  Journal}~{\bf 783},  67 (March 2014).

\bibitem{bkp2015}
{BICEP2/Keck and Planck Collaborations}, Ade, P. A.~R., Aghanim, N., Ahmed, Z.,
  Aikin, R.~W., Alexander, K.~D., Arnaud, M., Aumont, J., Baccigalupi, C.,
  Banday, A.~J., Barkats, D., Barreiro, R.~B., Bartlett, J.~G., Bartolo, N.,
  Battaner, E., Benabed, K., Beno\^{\i}t, A., Benoit-L\'evy, A., Benton, S.~J.,
  Bernard, J.-P., Bersanelli, M., Bielewicz, P., Bischoff, C.~A., Bock, J.~J.,
  Bonaldi, A., Bonavera, L., Bond, J.~R., Borrill, J., Bouchet, F.~R.,
  Boulanger, F., Brevik, J.~A., Bucher, M., Buder, I., Bullock, E., Burigana,
  C., Butler, R.~C., Buza, V., Calabrese, E., Cardoso, J.-F., Catalano, A.,
  Challinor, A., Chary, R.-R., Chiang, H.~C., Christensen, P.~R., Colombo, L.
  P.~L., Combet, C., Connors, J., Couchot, F., Coulais, A., Crill, B.~P.,
  Curto, A., Cuttaia, F., Danese, L., Davies, R.~D., Davis, R.~J.,
  de~Bernardis, P., de~Rosa, A., de~Zotti, G., Delabrouille, J., Delouis,
  J.-M., D\'esert, F.-X., Dickinson, C., Diego, J.~M., Dole, H., Donzelli, S.,
  Dor\'e, O., Douspis, M., Dowell, C.~D., Duband, L., Ducout, A., Dunkley, J.,
  Dupac, X., Dvorkin, C., Efstathiou, G., Elsner, F., En\ss{}lin, T.~A.,
  Eriksen, H.~K., Falgarone, E., Filippini, J.~P., Finelli, F., Fliescher, S.,
  Forni, O., Frailis, M., Fraisse, A.~A., Franceschi, E., Frejsel, A.,
  Galeotta, S., Galli, S., Ganga, K., Ghosh, T., Giard, M., Gjerl\o{}w, E.,
  Golwala, S.~R., Gonz\'alez-Nuevo, J., G\'orski, K.~M., Gratton, S., Gregorio,
  A., Gruppuso, A., Gudmundsson, J.~E., Halpern, M., Hansen, F.~K., Hanson, D.,
  Harrison, D.~L., Hasselfield, M., Helou, G., Henrot-Versill\'e, S., Herranz,
  D., Hildebrandt, S.~R., Hilton, G.~C., Hivon, E., Hobson, M., Holmes, W.~A.,
  Hovest, W., Hristov, V.~V., Huffenberger, K.~M., Hui, H., Hurier, G., Irwin,
  K.~D., Jaffe, A.~H., Jaffe, T.~R., Jewell, J., Jones, W.~C., Juvela, M.,
  Karakci, A., Karkare, K.~S., Kaufman, J.~P., Keating, B.~G., Kefeli, S.,
  Keih\"anen, E., Kernasovskiy, S.~A., Keskitalo, R., Kisner, T.~S., Kneissl,
  R., Knoche, J., Knox, L., Kovac, J.~M., Krachmalnicoff, N., Kunz, M., Kuo,
  C.~L., Kurki-Suonio, H., Lagache, G., L\"ahteenm\"aki, A., Lamarre, J.-M.,
  Lasenby, A., Lattanzi, M., Lawrence, C.~R., Leitch, E.~M., Leonardi, R.,
  Levrier, F., Lewis, A., Liguori, M., Lilje, P.~B., Linden-V\o{}rnle, M.,
  L\'opez-Caniego, M., Lubin, P.~M., Lueker, M., Mac\'{\i}as-P\'erez, J.~F.,
  Maffei, B., Maino, D., Mandolesi, N., Mangilli, A., Maris, M., Martin, P.~G.,
  Mart\'{\i}nez-Gonz\'alez, E., Masi, S., Mason, P., Matarrese, S., Megerian,
  K.~G., Meinhold, P.~R., Melchiorri, A., Mendes, L., Mennella, A., Migliaccio,
  M., Mitra, S., Miville-Desch\^enes, M.-A., Moneti, A., Montier, L., Morgante,
  G., Mortlock, D., Moss, A., Munshi, D., Murphy, J.~A., Naselsky, P., Nati,
  F., Natoli, P., Netterfield, C.~B., Nguyen, H.~T., N\o{}rgaard-Nielsen,
  H.~U., Noviello, F., Novikov, D., Novikov, I., O'Brient, R., Ogburn, R.~W.,
  Orlando, A., Pagano, L., Pajot, F., Paladini, R., Paoletti, D., Partridge,
  B., Pasian, F., Patanchon, G., Pearson, T.~J., Perdereau, O., Perotto, L.,
  Pettorino, V., Piacentini, F., Piat, M., Pietrobon, D., Plaszczynski, S.,
  Pointecouteau, E., Polenta, G., Ponthieu, N., Pratt, G.~W., Prunet, S.,
  Pryke, C., Puget, J.-L., Rachen, J.~P., Reach, W.~T., Rebolo, R., Reinecke,
  M., Remazeilles, M., Renault, C., Renzi, A., Richter, S., Ristorcelli, I.,
  Rocha, G., Rossetti, M., Roudier, G., Rowan-Robinson, M., Rubi\~no
  Mart\'{\i}n, J.~A., Rusholme, B., Sandri, M., Santos, D., Savelainen, M.,
  Savini, G., Schwarz, R., Scott, D., Seiffert, M.~D., Sheehy, C.~D., Spencer,
  L.~D., Staniszewski, Z.~K., Stolyarov, V., Sudiwala, R., Sunyaev, R., Sutton,
  D., Suur-Uski, A.-S., Sygnet, J.-F., Tauber, J.~A., Teply, G.~P., Terenzi,
  L., Thompson, K.~L., Toffolatti, L., Tolan, J.~E., Tomasi, M., Tristram, M.,
  Tucci, M., Turner, A.~D., Valenziano, L., Valiviita, J., Van~Tent, B.,
  Vibert, L., Vielva, P., Vieregg, A.~G., Villa, F., Wade, L.~A., Wandelt,
  B.~D., Watson, R., Weber, A.~C., Wehus, I.~K., White, M., White, S. D.~M.,
  Willmert, J., Wong, C.~L., Yoon, K.~W., Yvon, D., Zacchei, A., and Zonca, A.,
  ``{Joint Analysis of BICEP2/ \textit{Keck Array} and \textit{Planck} Data},''
  {\em Phys. Rev. Lett.}~{\bf 114},  101301 (March 2015).

\bibitem{ahmed2014}
{Ahmed}, Z., {Amiri}, M., {Benton}, S.~J., {Bock}, J.~J., {Bowens-Rubin}, R.,
  {Buder}, I., {Bullock}, E., {Connors}, J., {Filippini}, J.~P., {Grayson},
  J.~A., {Halpern}, M., {Hilton}, G.~C., {Hristov}, V.~V., {Hui}, H., {Irwin},
  K.~D., {Kang}, J., {Karkare}, K.~S., {Karpel}, E., {Kovac}, J.~M., {Kuo},
  C.~L., {Netterfield}, C.~B., {Nguyen}, H.~T., {O'Brient}, R., {Ogburn},
  R.~W., {Pryke}, C., {Reintsema}, C.~D., {Richter}, S., {Thompson}, K.~L.,
  {Turner}, A.~D., {Vieregg}, A.~G., {Wu}, W.~L.~K., and {Yoon}, K.~W.,
  ``{BICEP3: a 95 GHz refracting telescope for degree-scale CMB
  polarization},'' in [{\em Millimeter, Submillimeter, and Far-Infrared
  Detectors and Instrumentation for Astronomy
  VII}{\nolinebreak\hspace{0.1em}]},  {\em Proc. SPIE} {\bf 9153},  91531N
  (August 2014).

\bibitem{thesis:wu}
Wu, W. L.~K., {\em {BICEP3 and CMB-S4: current and future CMB polarization
  experiments to probe fundamental physics}}, PhD thesis, Stanford University
  (2015).

\bibitem{dets2015}
Ade, P. A.~R., Aikin, R.~W., Amiri, M., Barkats, D., Benton, S.~J., Bischoff,
  C.~A., Bock, J.~J., Bonetti, J.~A., Brevik, J.~A., Buder, I., Bullock, E.,
  Chattopadhyay, G., Davis, G., Day, P.~K., Dowell, C.~D., Duband, L.,
  Filippini, J.~P., Fliescher, S., Golwala, S.~R., Halpern, M., Hasselfield,
  M., Hildebrandt, S.~R., Hilton, G.~C., Hristov, V., Hui, H., Irwin, K.~D.,
  Jones, W.~C., Karkare, K.~S., Kaufman, J.~P., Keating, B.~G., Kefeli, S.,
  Kernasovskiy, S.~A., Kovac, J.~M., Kuo, C.~L., LeDuc, H.~G., Leitch, E.~M.,
  Llombart, N., Lueker, M., Mason, P., Megerian, K., Moncelsi, L., Netterfield,
  C.~B., Nguyen, H.~T., O’Brient, R., IV, R. W.~O., Orlando, A., Pryke, C.,
  Rahlin, A.~S., Reintsema, C.~D., Richter, S., Runyan, M.~C., Schwarz, R.,
  Sheehy, C.~D., Staniszewski, Z.~K., Sudiwala, R.~V., Teply, G.~P., Tolan,
  J.~E., Trangsrud, A., Tucker, R.~S., Turner, A.~D., Vieregg, A.~G., Weber,
  A., Wiebe, D.~V., Wilson, P., Wong, C.~L., Yoon, K.~W., Zmuidzinas, J., and
  {for the BICEP2, Keck Array, and Spider Collaborations}, ``{Antenna-coupled
  TES Bolometers Used in BICEP2, Keck Array, and Spider},'' {\em The
  Astrophysical Journal}~{\bf 812},  176 (October 2015).

\bibitem{obrient2012}
{O'Brient}, R., {Ade}, P.~A.~R., {Ahmed}, Z., {Aikin}, R.~W., {Amiri}, M.,
  {Benton}, S., {Bischoff}, C., {Bock}, J.~J., {Bonetti}, J.~A., {Brevik},
  J.~A., {Burger}, B., {Davis}, G., {Day}, P., {Dowell}, C.~D., {Duband}, L.,
  {Filippini}, J.~P., {Fliescher}, S., {Golwala}, S.~R., {Grayson}, J.,
  {Halpern}, M., {Hasselfield}, M., {Hilton}, G., {Hristov}, V.~V., {Hui}, H.,
  {Irwin}, K., {Kernasovskiy}, S., {Kovac}, J.~M., {Kuo}, C.~L., {Leitch}, E.,
  {Lueker}, M., {Megerian}, K., {Moncelsi}, L., {Netterfield}, C.~B., {Nguyen},
  H.~T., {Ogburn}, R.~W., {Pryke}, C.~L., {Reintsema}, C., {Ruhl}, J.~E.,
  {Runyan}, M.~C., {Schwarz}, R., {Sheehy}, C.~D., {Staniszewski}, Z.,
  {Sudiwala}, R., {Teply}, G., {Tolan}, J.~E., {Turner}, A.~D., {Tucker},
  R.~S., {Vieregg}, A., {Wiebe}, D.~V., {Wilson}, P., {Wong}, C.~L., {Wu},
  W.~L.~K., and {Yoon}, K.~W., ``{Antenna-coupled TES bolometers for the Keck
  array, Spider, and Polar-1},'' in [{\em Millimeter, Submillimeter, and
  Far-Infrared Detectors and Instrumentation for Astronomy
  VI}{\nolinebreak\hspace{0.1em}]},  {\em Proc. SPIE} {\bf 8452},  84521G
  (September 2012).

\bibitem{spie:howard}
{Hui, H. et~al.}, ``{Second year detectors performance of BICEP3},'' in [{\em
  these proceedings}{\nolinebreak\hspace{0.1em}]},  {\em Proc. SPIE} (2016).

\bibitem{dekorte2003}
de~Korte, P. A.~J., Beyer, J., Deiker, S., Hilton, G.~C., Irwin, K.~D.,
  MacIntosh, M., Nam, S.~W., Reintsema, C.~D., Vale, L.~R., and Huber, M.~E.,
  ``Time-division superconducting quantum interference device multiplexer for
  transition-edge sensors,'' {\em Review of Scientific Instruments}~{\bf
  74}(8),  3807--3815 (2003).

\bibitem{mce2008}
{Battistelli}, E.~S., {Amiri}, M., {Burger}, B., {Halpern}, M., {Knotek}, S.,
  {Ellis}, M., {Gao}, X., {Kelly}, D., {Macintosh}, M., {Irwin}, K., and
  {Reintsema}, C., ``{Functional Description of Read-out Electronics for
  Time-Domain Multiplexed Bolometers for Millimeter and Sub-millimeter
  Astronomy},'' {\em Journal of Low Temperature Physics}~{\bf 151},  908--914
  (May 2008).

\bibitem{mesh2014}
Ahmed, Z., Grayson, J.~A., Thompson, K.~L., Kuo, C.-L., Brooks, G., and
  Pothoven, T., ``{Large-Area Reflective Infrared Filters for Millimeter/Sub-mm
  Telescopes},'' {\em Journal of Low Temperature Physics}~{\bf 176}(5),
  835--840 (2014).

\bibitem{ade2006}
Ade, P. A.~R., Pisano, G., Tucker, C., and Weaver, S., ``A review of metal mesh
  filters,'' {\em Proc. SPIE}~{\bf 6275},  62750U--62750U--15 (2006).

\bibitem{wu2015}
Wu, W. L.~K., Ade, P. A.~R., Ahmed, Z., Alexander, K.~D., Amiri, M., Barkats,
  D., Benton, S.~J., Bischoff, C.~A., Bock, J.~J., Bowens-Rubin, R., Buder, I.,
  Bullock, E., Buza, V., Connors, J.~A., Filippini, J.~P., Fliescher, S.,
  Grayson, J.~A., Halpern, M., Harrison, S., Hilton, G.~C., Hristov, V.~V.,
  Hui, H., Irwin, K.~D., Kang, J., Karkare, K.~S., Karpel, E., Kefeli, S.,
  Kernasovskiy, S.~A., Kovac, J.~M., Kuo, C.~L., Megerian, K.~G., Netterfield,
  C.~B., Nguyen, H.~T., O'Brient, R., Ogburn, R.~W., Pryke, C., Reintsema,
  C.~D., Richter, S., Sorensen, C., Staniszewski, Z.~K., Steinbach, B.,
  Sudiwala, R.~V., Teply, G.~P., Thompson, K.~L., Tolan, J.~E., Tucker, C.~E.,
  Turner, A.~D., Vieregg, A.~G., Weber, A.~C., Wiebe, D.~V., Willmert, J., and
  Yoon, K.~W., ``{Initial Performance of BICEP3: A Degree Angular Scale 95 GHz
  Band Polarimeter},'' {\em Journal of Low Temperature Physics}  (December
  2015).

\bibitem{karkare2014}
Karkare, K.~S., Ade, P. A.~R., Ahmed, Z., Aikin, R.~W., Alexander, K.~D.,
  Amiri, M., Barkats, D., Benton, S.~J., Bischoff, C.~A., Bock, J.~J., Bonetti,
  J.~A., Brevik, J.~A., Buder, I., Bullock, E.~W., Burger, B., Connors, J.,
  Crill, B.~P., Davis, G., Dowell, C.~D., Duband, L., Filippini, J.~P.,
  Fliescher, S.~T., Golwala, S.~R., Gordon, M.~S., Grayson, J.~A., Halpern, M.,
  Hasselfield, M., Hildebrandt, S.~R., Hilton, G.~C., Hristov, V.~V., Hui, H.,
  Irwin, K.~D., Kang, J.~H., Karpel, E., Kefeli, S., Kernasovskiy, S.~A.,
  Kovac, J.~M., Kuo, C.~L., Leitch, E.~M., Lueker, M., Mason, P., Megerian,
  K.~G., Netterfield, C.~B., Nguyen, H.~T., O'Brient, R., Ogburn, R.~W., Pryke,
  C.~L., Reintsema, C.~D., Richter, S., Schwarz, R., Sheehy, C.~D.,
  Staniszewski, Z.~K., Sudiwala, R.~V., Teply, G.~P., Thompson, K.~L., Tolan,
  J.~E., Turner, A.~D., Vieregg, A., Weber, A., Wong, C.~L., Wu, W. L.~K., and
  Yoon, K.~W., ``{ Keck array and BICEP3: spectral characterization of 5000+
  detectors },'' in [{\em Millimeter, Submillimeter, and Far-Infrared Detectors
  and Instrumentation for Astronomy VII}{\nolinebreak\hspace{0.1em}]},  {\em
  Proc. SPIE}~{\bf 9153},  91533B--91533B--11 (August 2014).

\bibitem{spie:kirit}
{Karkare, K.~S. et~al.}, ``{Optical characterization of the BICEP3 CMB
  polarimeter at the South Pole},'' in [{\em these
  proceedings}{\nolinebreak\hspace{0.1em}]},  {\em Proc. SPIE} (2016).

\end{thebibliography}
